# Dimensionless scaling of heat-release-induced planar shock waves in near-critical $CO_2$

Mario Tindaro Migliorino[*] and Carlo Scalo[†]

*Purdue University, West Lafayette, IN 47907, USA*

We performed highly resolved one-dimensional fully compressible Navier-Stokes simulations of heat-release-induced compression waves in near-critical $CO_2$. The computational setup, inspired by the experimental setup of Miura et al., *Phys. Rev. E,* **2006**, is composed of a closed inviscid (one-dimensional) duct of 150 $\mu$m in length with adiabatic hard ends filled with $CO_2$ at initial pressures $p_0^* = 1.00, 1.04, 1.40\ p_c^*$, where $p_c^* = 7.3773$ MPa is the critical pressure of $CO_2$ and the superscript ($^*$) denotes dimensional quantities. The corresponding initial temperature values are taken along the pseudo-boiling line. Thermodynamic and transport properties of $CO_2$ in near-critical conditions are modeled via the Peng-Robinson equation of state and Chung's Method. A heat source is applied at a distance of 50 $\mu$m from one end, with heat release intensities per unit area, $\Omega^*$, spanning the range $10^3 - 10^{11}$ W/m$^2$, generating isentropic compression waves for $\Omega^* < 10^9$ W/m$^2$. For higher heat-release rates such compressions are coalescent with distinct shock-like features (e.g. non-isentropicity and propagation Mach numbers measurably greater than unity) and a non-uniform post-shock state is present due to the strong thermodynamic nonlinearities. The resulting compression wave intensities have been collapsed via the thermal expansion coefficient, highly variable in near-critical fluids, used as one of the scaling parameters for the reference energy. The proposed scaling applies to isentropic thermoacoustic waves as well as shock waves up to shock strength $\Pi = 2$. Long-term time integration reveals resonance behavior of the compression waves, raising the mean pressure and temperature at every (near-acoustic) resonance cycle. This phenomenon is known in the literature as "Piston Effect". When the heat injection is halted, expansion waves are generated, which counteract the compression waves leaving conduction as the only thermal relaxation process. In the long term evolution, the decay in amplitude of the resonating waves observed in the experiments is qualitatively reproduced by using isothermal boundary conditions. Future efforts will focus on developing appropriate wall-impedance conditions for near-critical fluids.

## I. Introduction

### I.A. Background

WHEN a compressible fluid is thermally perturbed, a mechanical response is generated in the form of waves,[16,30] which are referred to by several authors as thermoacoustic waves. Depending on the intensity of the heat release, such waves can range from isentropic compressions to shock waves.[8] The amplitude of the thus-obtained thermoacoustic waves is also expected to strongly depend on the initial unperturbed state of the fluid. Moreover, very different responses can be observed in the case of ideal gases as opposed to supercritical fluids. In particular, near-critical fluids are more sensitive to heat release, given their peculiar thermodynamic features.[5,32] These include a sudden drop in density when temperature is raised: heavy fluid (pseudo liquid) transitions to lighter fluid (pseudo gas) following a "pseudo-boiling"[1] process, during which thermal diffusivity becomes evanescent due to the peak in isobaric heat capacity. Because of this, heat transport in enclosed single-component near-critical fluids, thought to be exclusively linked to molecular diffusion, was expected to be very slow compared to acoustic time scales. However, experiments by Nitsche et al.[23] and Klein et al.,[17] carried out under reduced gravity, have shown that in an enclosed near-critical fluid small temperature changes propagate much faster than diffusion-only processes. Several groups[2,24,33] reproduced this phenomenon, eventually named Piston Effect (PE).[33] According to the PE theory, thermoacoustic waves generated by heat addition can heat up a closed volume sample on the acoustic time scales. This occurs via resonance of compression waves raising mean temperature and pressure at every cycle, as shown by experiments of heat injection into closed gas columns.[3,11,13] A large number of numerical studies[9,11,12,22,28,29,32,34] has also been carried out, relying on the van der Waals equation of state, or interpolating thermodynamic quantities from the NIST[19] database, or implementing linearized equations of state.

---

[*]Ph.D. Student, Department of Mechanical Engineering, AIAA Member, email: migliom@purdue.edu
[†]Assistant Professor, Department of Mechanical Engineering, AIAA Member, email: scalo@purdue.edu

### I.B. Research aims

The goal of the present study is to investigate scaling properties of thermoacoustic waves in near-critical fluids and also attempt to lay the theoretical foundation for acoustics in supercritical fluids. The current work is inspired by the canonical investigation of Miura et al.,[21] where measurements of density variations of about $10^{-7}$ g/cm$^3$ in amplitude from isentropic thermoacoustic waves were performed with an ultrasensitive interferometer. This experiment has been also modeled with asymptotic matching techniques.[4]

We propose a gas dynamic approach in the interpretation of the PE, viewing all the heat-release-generated waves as shock waves, regardless of their intensity. In order to accurately reproduce heat-release-generated compression waves, we performed one-dimensional high-fidelity numerical simulations solving the fully compressible Navier-Stokes equations. We rely on the Peng-Robinson[26] equation of state (EoS), which has the advantage of making analytical thermodynamic derivatives available, guaranteeing thermodynamic consistency, and aiding the development of low order models.

We propose in this paper a general scaling of isentropic thermoacoustic wave amplitudes that stems out of a dimensional analysis of the compressible Navier-Stokes equations, and differentials of thermodynamic quantities, with a generic equation of state. We extend this scaling, derived from isentropic principles, to heat-release-generated shock waves employing the isentropic exponent[14] as an effective ratio of specific heats $\widetilde{\gamma}$. Finally, we show calculations capturing several resonance cycles of isentropic thermoacoustic waves reproducing the experiments of Miura et al.[21]

This paper is organized as follows: the governing equations in dimensional form and the proposed thermoacoustic scaling, leading to their dimensionless form, are shown in section II. The problem formulation is described in section III, which includes the numerical and computational setup. Finally, results are proposed with a detailed discussion in section IV.

## II. Thermoacoustically scaled governing equations

The one-dimensional compressible Navier-Stokes equations are reported in this section. We first show the governing equations in dimensional form, then report the set of reference variables at the core of the proposed thermoacoustic scaling. Finally, we present the governing equations made dimensionless accordingly.

### II.A. Dimensional governing equations

The complete set of single-phase, one-dimensional compressible Navier-Stokes equations representing the conservation of mass, momentum and total energy, read, respectively:

$$\frac{\partial \rho^*}{\partial t^*} + \frac{\partial \rho^* u^*}{\partial x^*} = 0, \tag{1}$$

$$\frac{\partial \rho^* u^*}{\partial t^*} + \frac{\partial \rho^* u^{*2}}{\partial x^*} = -\frac{\partial p^*}{\partial x^*} + \frac{\partial \tau^*}{\partial x^*}, \tag{2}$$

$$\frac{\partial \rho^* E^*}{\partial t^*} + \frac{\partial \rho^* E^* u^*}{\partial x^*} = -\frac{\partial p^* u^*}{\partial x^*} + \frac{\partial u^* \tau^*}{\partial x^*} - \frac{\partial q^*}{\partial x^*} + \dot{Q}^*, \tag{3}$$

where the superscript (*) indicates dimensional quantities, $t^*$ is time, $x^*$ is the spatial coordinate, $u^*$ the velocity, $\rho^*$ is the density, and $p^*, T^*$ are the thermodynamic pressure and temperature. The specific total energy $E^* = e^* + u^{*2}/2$ is the sum of internal energy $e^*$ and kinetic energy $u^{*2}/2$. The Newtonian viscous stresses are expressed according to Stokes's hypothesis,

$$\tau^* = \frac{4}{3}\mu^* \frac{\partial u^*}{\partial x^*}, \tag{4}$$

and the heat flux is modeled with Fourier heat conduction,

$$q^* = -k^* \frac{\partial T^*}{\partial x^*}, \tag{5}$$

where $\mu^* = \mu^*(\rho^*, T^*)$ is the dynamic viscosity and $k^* = k^*(\rho^*, T^*)$ is the thermal conductivity. Finally, $\dot{Q}^*$ is the volumetric heat release rate expressed in W/m³. No other source terms are considered. Equations (1),(2),(3), with the thermodynamic closure given by the equation of state discussed in section III.A, and with given initial and boundary conditions, can be numerically solved.

Energy conservation will hereafter be recast in forms suitable for deriving the proposed novel thermoacoustic scaling. From eq. (3), the definition of total energy and eq.s (1) and (2), one can derive the transport equation for internal energy:

$$\rho^* \frac{De^*}{Dt^*} = -p^* \frac{\partial u^*}{\partial x^*} + \Phi^* - \frac{\partial q^*}{\partial x^*} + \dot{Q}^*, \tag{6}$$

where $D/Dt^*$ is the material derivative and $\Phi^*$ the viscous dissipation function, given by

$$\Phi^* = \tau^* \frac{\partial u^*}{\partial x^*}. \tag{7}$$

Eq. (6) allows to write the evolution equation for enthalpy $h^* = e^* + p^*/\rho^*$ (not reported). The Gibbs relation

$$\rho^* T^* ds^* = \rho^* de^* - \frac{p^*}{\rho^*} d\rho^*, \tag{8}$$

where $s^*$ is the specific entropy, together with eq. (6) allows to obtain the entropy transport equation,

$$\rho^* T^* \frac{Ds^*}{Dt^*} = \Phi^* - \frac{\partial q^*}{\partial x^*} + \dot{Q}^*. \tag{9}$$

Eq. (9) shows that the entropy will vary due to viscous dissipation, heat diffusion, and heat release. The thermodynamic expression of the total differential of the internal energy for a compressible fluid, with a generic equation of state,

$$p^* = p^*(\rho^*, T^*), \tag{10}$$

reads

$$de^* = c_v^* dT^* - \left(\frac{c_v^*}{\rho^* \alpha_p^*}(\gamma - 1) - \frac{p^*}{\rho^{*2}}\right) d\rho^*, \tag{11}$$

with

$$\frac{a^{*2} T^* \alpha_p^{*2}}{c_p^*} = \gamma - 1 \tag{12}$$

where $c_p^*$ and $c_v^* = c_p^*/\gamma$ are the specific isobaric and isochoric heat capacities, respectively, and $a^*$ is the speed of sound. *Note that the term in brackets on the right hand side of eq. (11) is zero for an ideal gas.* When deriving eq. (11), the thermal expansion coefficient $\alpha_p^*$

$$\alpha_p^* = -\frac{1}{\rho^*} \frac{\partial \rho^*}{\partial T^*}\bigg|_{p^*} \tag{13}$$

had to be introduced. As shown later, this term is the key to the proposed thermoacoustic scaling.

Eq. (6) can be combined with eq. (11) to get the temperature evolution equation

$$\rho^* c_v^* \frac{DT^*}{Dt^*} = -\frac{\rho^* c_v^*}{\alpha_p^*}(\gamma - 1)\frac{\partial u^*}{\partial x^*} + \Phi^* - \frac{\partial q^*}{\partial x^*} + \dot{Q}^*. \tag{14}$$

Differentiating eq. (10) yields

$$\gamma \frac{dp^*}{\rho^* a^{*2}} = \frac{d\rho^*}{\rho^*} + \alpha_p^* dT^*, \tag{15}$$

which combined with eq. (14) and (1) returns the evolution equation for pressure:

$$\frac{Dp^*}{Dt^*} = -\rho^* a^{*2} \frac{\partial u^*}{\partial x^*} + \frac{\alpha_p^* a^{*2}}{c_p^*} \left( \Phi^* - \frac{\partial q^*}{\partial x^*} + \dot{Q}^* \right). \tag{16}$$

### II.B. Proposed thermoacoustic scaling

The first terms on the right hand side of eq. (14) and eq. (16) govern the inviscid isentropic response of the fluid. This informs how isentropic perturbations should be scaled: the pressure should scale with $\rho^* a^{*2}$, consistently with traditional linear acoustic theory, while changes in temperature should scale as $(\gamma - 1)/\alpha_p^*$. This would allow to obtain, in the corresponding dimensionless equations, a unitary multiplicative coefficient of the material derivative of pressure and temperature, and of the velocity divergence.

This heuristic inspection has inspired a new thermoacoustic scaling, outlined in the following. Scaling parameters are hereafter indicated with a subscript "0", which will coincide with the reference base or initial state of the heat-release numerical experiments. Scaling of kinematic quantities reads

$$x = \frac{x^*}{l_0^*}, \qquad t = \frac{t^*}{l_0^*/a_0^*}, \qquad u = \frac{u^*}{a_0^*}, \tag{17}$$

where $l_0^*$ is an arbitrary reference length scale. The non-dimensionalization of thermodynamic variables reads

$$\rho = \frac{\rho^*}{\rho_0^*}, \qquad T = \frac{T^*}{(\gamma_0 - 1)/\alpha_{p_0}^*}, \qquad p = \frac{p^*}{\rho_0^* a_0^{*2}}, \tag{18}$$

in accordance with the aforementioned considerations. The dimensionless energy-related variables are

$$E = \frac{E^*}{c_{p_0}^*/\alpha_{p_0}^*}, \qquad e = \frac{e^*}{c_{p_0}^*/\alpha_{p_0}^*}, \qquad h = \frac{h^*}{c_{p_0}^*/\alpha_{p_0}^*}, \tag{19}$$

where the reference energy is chosen as $c_{p_0}^*/\alpha_{p_0}^*$. The dimensionless specific thermal coefficients are:

$$c_p = \frac{c_p^*}{c_{p_0}^*}, \qquad c_v = \frac{c_v^*}{c_{p_0}^*}. \tag{20}$$

Remaining variables are normalized as follows

$$a = \frac{a^*}{a_0^*}, \qquad \alpha_p = \frac{\alpha_p^*}{\alpha_{p_0}^*}, \qquad \mu = \frac{\mu^*}{\mu_0^*}, \qquad k = \frac{k^*}{k_0^*}. \tag{21}$$

The energy source term, as shown in section IV.C, should be normalized as

$$\dot{Q} = \frac{\dot{Q}^*}{(\rho_0^* a_0^* c_{p_0}^*/\alpha_{p_0}^*)/l_0^*}, \tag{22}$$

allowing a self-similar collapse of the intensity of heat-release-induced compression waves.

### II.C. Thermoacoustically scaled governing equations

Now we show the governing equations, made dimensionless based on scaling presented in section II.B, starting with mass, momentum and total energy:

$$\frac{\partial \rho}{\partial t} + \frac{\partial \rho u}{\partial x} = 0, \tag{23}$$

$$\frac{\partial \rho u}{\partial t} + \frac{\partial \rho u^2}{\partial x} = -\frac{\partial p}{\partial x} + \frac{1}{\text{Re}} \frac{\partial \tau}{\partial x}, \tag{24}$$

$$\frac{\partial \rho E}{\partial t} + \frac{\partial \rho E u}{\partial x} = -\text{Ec}\frac{\partial p u}{\partial x} + \frac{\text{Ec}}{\text{Re}}\frac{\partial u \tau}{\partial x} - \frac{\gamma_0 - 1}{\text{Pe}}\frac{\partial q}{\partial x} + \dot{Q}, \qquad (25)$$

where the sonic Reynolds, Péclet, Eckert, and Prandtl numbers are, respectively:

$$\text{Re} = \frac{\rho_0^* a_0^* l_0^*}{\mu_0^*}, \quad \text{Pe} = \text{RePr}, \quad \text{Ec} = \frac{a_0^{*2}}{c_{p_0}^*/\alpha_{p_0}^*} \qquad \text{Pr} = \frac{\mu_0^* c_{p_0}^*}{k_0^*}. \qquad (26)$$

The dimensionless evolution equations for temperature and pressure read, respectively,

$$\rho c_v \frac{DT}{Dt} = -\frac{\rho c_v}{\alpha_p}\frac{\gamma - 1}{\gamma_0 - 1}\frac{\partial u}{\partial x} + \frac{\text{Ec}}{(\gamma_0 - 1)\text{Re}}\Phi - \frac{1}{\text{Pe}}\frac{\partial q}{\partial x} + \frac{\dot{Q}}{\gamma_0 - 1}, \qquad (27)$$

and

$$\frac{Dp}{Dt} = -\rho a^2 \frac{\partial u}{\partial x} + \frac{\alpha_p a^2}{c_p}\left(\frac{\text{Ec}}{\text{Re}}\Phi - \frac{\gamma_0 - 1}{\text{Pe}}\frac{\partial q}{\partial x} + \dot{Q}\right). \qquad (28)$$

Dimensionless internal energy and entropy equations are here omitted for brevity.

By inspecting equations (27) and (28) it can be noted that the proposed thermoacoustic scaling results in isentropic temperature perturbations not only being of the same order of, but also equal to pressure perturbations. In fact, assuming isentropic flow yields

$$\frac{DT}{Dt} = -\frac{1}{\alpha_p}\frac{\gamma - 1}{\gamma_0 - 1}\frac{\partial u}{\partial x}, \qquad \frac{Dp}{Dt} = -\rho a^2 \frac{\partial u}{\partial x}, \qquad (29)$$

which linearized about the (dimensionless) base state $\gamma = \gamma_0, \alpha_{p_0} = \rho_0 = a_0 = 1$, becomes

$$\frac{D[T]}{Dt} = \frac{D[p]}{Dt} = -\frac{\partial [u]}{\partial x}, \qquad (30)$$

where square brackets indicate fluctuations or jumps about the base state denoted by subscript "0". The dynamics of fluctuations associated with compression waves will be analyzed in section IV.

## III. Problem formulation

### III.A. Fluid model

The Peng-Robinson[26] equation of state (PR EoS) is chosen as real-fluid model because of its thermodynamic consistency, simplicity, and accuracy for the parameter space explored in this study. The PR EoS relates pressure $p^*$, temperature $T^*$ and specific volume $v^* = 1/\rho^*$ via

$$p^* = \frac{R_u^* T^*}{v_m^* - b_m^*} - \frac{\alpha_m^*}{v_m^{*2} + 2v_m^* b_m^* - b_m^{*2}} \qquad (31)$$

where $v_m^* = M_m^* v^*$ is the molar volume, $M_m^*$ is the molar mass of the substance, and $R^* = R_u^*/M_m^*$ where $R_u^* = 8.314472$ J·mol$^{-1}$K$^{-1}$ is the universal gas constant. The other coefficients in eq. (31) are

$$\alpha_m^*(T^*) = 0.45724\frac{R_u^{*2} T_c^{*2}}{p_c^*}\alpha(T^*), \qquad \alpha(T^*) = \left[1 + \beta\left(1 - \sqrt{T^*/T_c^*}\right)\right]^2, \qquad (32)$$

$$b_m^* = 0.07780\frac{R_u^* T_c^*}{p_c^*}, \qquad \beta = 0.37464 + 1.54226\omega - 0.26992\omega^2 \qquad (33)$$

where $\omega$ is the acentric factor of the substance. The subscript "c" indicates thermodynamic quantities at the critical point. Table 1 reports values of the aforementioned fluid model parameters relative to carbon dioxide.[19,27] Notice that an acentric factor of 0.225, rather than the most recent measured value[19] of 0.22394, should be used, as pointed out by Poling *et al.* [27, p. 2.26], since the former was the one originally used when deriving the correlations in (32) and (33).

All of the thermodynamic derivatives required by the current study, here omitted for brevity, can be computed directly, retaining full thermodynamic consistency, from eq. (31). In particular, they can be used to compute any generic thermodynamic quantity with the thermodynamic departure functions.[27] With this approach, a first contribution at constant volume is given by assuming a thermodynamic transformation, following ideal gas law, from $T^* = 0$ to $T^*$, reaching the state $(T^*, v^{0*})$ where $v^{0*} = R^* T^*/p^{0*}$ and $p^{0*} = 1$ bar.[20] Consequently, a volume integration, departing from the thus-derived ideal gas reference state, is performed to the final state $(T^*, v^*)$. This strategy can be applied to compute, for example, the specific isochoric, $c_v^*(T^*, v^*)$, and isobaric, $c_p^*(T^*, v^*)$, heat capacities by merely correcting their ideal gas counterparts, obtained from the polynomial fit given in the appendix of Poling et. al.[27]

| fluid | $T_c^*$(K) | $p_c^*$(MPa) | $\rho_c^*$(kg/m$^3$) | $v_{m_c}^*$(cm$^3$/mol) | $M_m^*$(g/mol) | $R^*$(J/kgK) | $\omega$ |
|---|---|---|---|---|---|---|---|
| CO$_2$ | 304.1282 | 7.3773 | 467.6 | 94.1189 | 44.01 | 188.92 | 0.225 |

**Table 1. Fluid properties relative to carbon dioxide required by the PR EoS, with the exception of $\rho_c^*$ and $v_{m_c}^*$.**

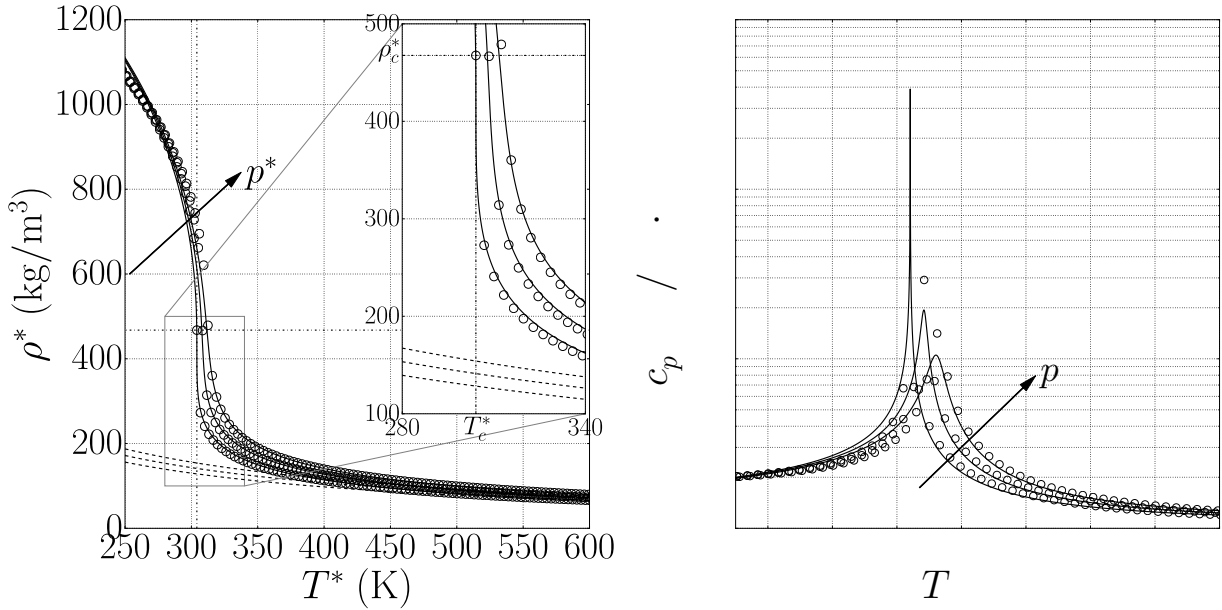

**Figure 1. Comparison between the Peng-Robinson equation of state (solid lines) and data from the NIST database (circles) for CO$_2$ at $p^* = 1.01, 1.10, 1.20\ p_c^*$. Density $\rho^* = \rho^*(T^*, p^*)$ on the left, and isobaric specific heat $c_p^*(T^*, p^*)$ on the right, are shown as a function of temperature. On the left, density as given by ideal gas law for the same pressure levels is shown with dashed lines.**

The PR EoS is in acceptable agreement with data from the NIST database[19] (figure 1) for CO$_2$ at the three selected pressures of $p^* = 1.01, 1.10, 1.20\ p_c^*$ and for temperatures higher than $T_c^*$. Figure 1 shows, on the left, the rapid drop in density approaching the critical point and how, for high temperatures, CO$_2$ behaves like an ideal gas. On the right, a very high thermal capacity in the near-critical region is shown. Dynamic viscosity and thermal conductivity are estimated via Chung's method,[7,27] also in fair agreement with NIST (figure 2) for CO$_2$.

In the near-critical region, the PR EoS fails to capture the experimental value of the critical density, being such equation tuned to the correct values of critical pressures and temperatures only, more easily measurable than the critical volume.[10] For carbon dioxide, in fact, the critical density predicted by PR EoS is 400 kg/m$^3$, which differs significantly from the experimental value $\rho_c^* = 467.6$ kg/m$^3$ (table 1).

Even if density corrections are available for cubic equations of state,[20] in this work the original version of the PR EoS is used. In fact the aim of this paper is not to reproduce exactly experimental measurements or quantitatively capture the near-critical behavior of CO$_2$, but rather to perform theoretical numerical simulations, inspired by experimental investigations, serving as a test bed for the proposed thermoacoustic scaling. Trying to modify the PR EoS to match critical densities would introduce unnecessary complications at this stage and will be considered in future work.

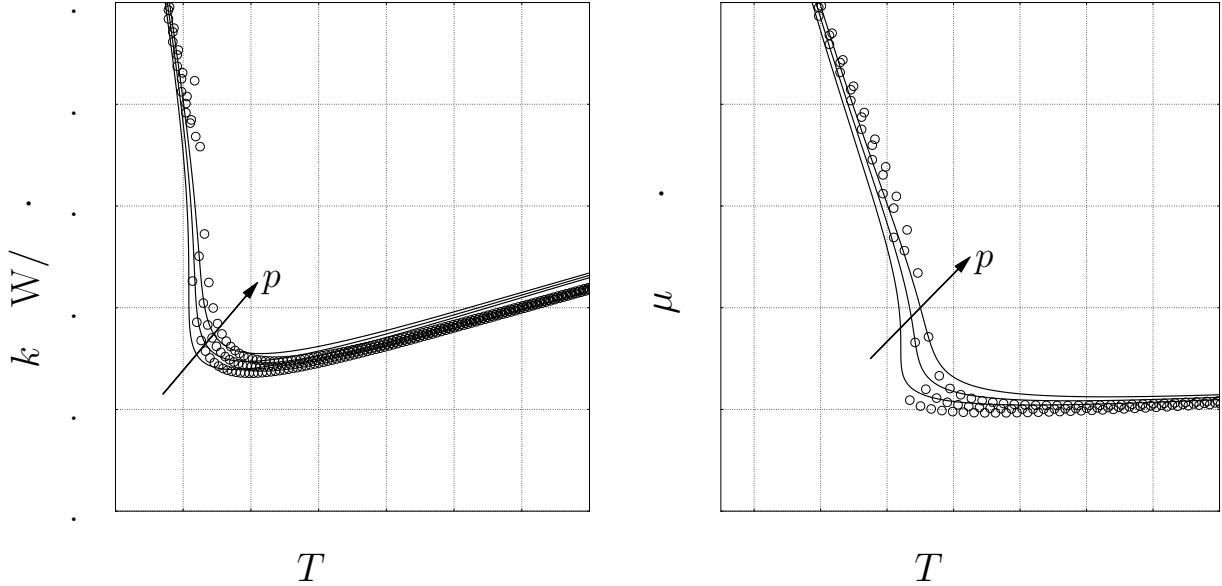

Figure 2. Comparison between Chung's method (solid lines) and data from the NIST database (circles) for $CO_2$ at $p^* = 1.01, 1.10, 1.20\ p_c^*$. Thermal conductivity, $k^* = k^*(T^*, p^*)$, on the left, and dynamic viscosity $\mu^* = \mu^*(T^*, p^*)$, on the right, are shown versus temperature.

### III.B. Numerical setup

The fully compressible, highly parallelized Navier-Stokes solver, *Hybrid*, developed by Larsson,[18] is used for the simulations. The code relies on a high-order-accurate spatial discretization. Time advancement is performed with a fourth order Runge-Kutta scheme. The CFL number is kept at 0.1 in all cases, resulting in a time step $\Delta t^*$ of the order of 0.01 ns. The high-order spatial reconstruction does not allow the presence of underresolved gradients without spurious numerical oscillations, unless shock capturing is active. In this work, however, the spatial resolution adopted is sufficiently high to resolve all gradients in the flow preventing the appearance of spurious numerical oscillations, therefore resolving and not capturing the shocks.

The code solves for a single-phase flow, an acceptable approach since surface tension is negligible for supercritical fluids.

### III.C. Computational setup

#### III.C.1. Functional form for heat release

The computational setup chosen for the present study is illustrated in figure 3 and is inspired from the experiment of Miura *et al.*[21] It consists of a duct of length $L^* = 150\ \mu$m confined by adiabatic or isothermal hard ends. Heat is applied at $x^* = 0$ to the fluid with a spatial and temporal distribution given by

$$\dot{Q}^*(x^*, t^*) = \begin{cases} \Omega^* f^*(x^*) & \text{if } 0 \leq t^* \leq t_f^* \\ 0 & \text{if } t^* > t_f^* \end{cases} \quad (34)$$

where $\Omega^*$ (energy release per unit area per unit time) is a tunable parameter determining the strength of the heat forcing. A step-like law is chosen in time: heat injection is active until $t_f^* = 0.1$ms, after which it is set to zero. The spatial distribution is determined by

$$f^*(x^*) = \frac{1}{\sigma^* \sqrt{2\pi}} \exp\left(-\frac{x^{*2}}{2\sigma^{*2}}\right) \quad (35)$$

which has dimensions of inverse of a length.

The heat source term has therefore the form

$$\dot{Q}^*(x^*, t^*) = f^*(x^*) \Omega^*(t^*). \quad (36)$$

7 of 16

American Institute of Aeronautics and Astronautics

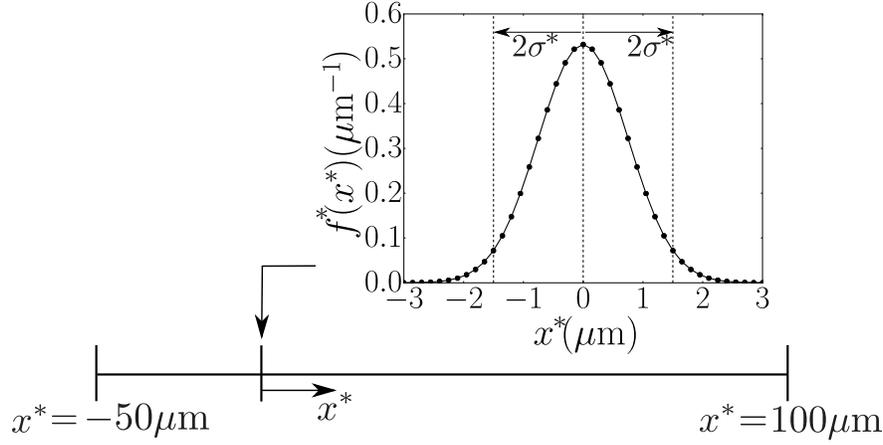

Figure 3. Computational setup and spatial distribution of heat injection.

The function $f^*(x^*)$ is a Gaussian distribution with unitary integral approximating a zero-thickness theoretical planar heat source. The volume integral intensity of the heat injection is equal to $\Omega^*$. The thickness of the heat release distribution, $4\sigma^*$, is chosen to be equal to 3 $\mu$m,[21] which results in $\sigma^* = 0.75$ $\mu$m, which is very small compared to the length of the tube but comparable to the obtained shock thicknesses. The very small thickness of the heated region could raise questions on the validity of the continuum hypothesis. The Knudsen number relevant to the present problem is:

$$\text{Kn} = \frac{\lambda^*}{\sigma^*} \tag{37}$$

where $\lambda^*$ is the mean free path of a Boltzmann gas, given by:

$$\lambda^* = \frac{k_B^* T^*}{\sqrt{2}\pi d^{*2} p^*} \tag{38}$$

where $k_B^* = 1.38064852 \cdot 10^{-23}$ J·K$^{-1}$ is the Boltzmann constant and $d^* = 3.3 \cdot 10^{-10}$ m is the kinetic diameter of $CO_2$.[15] At pressure and temperature $p^*/p_c^* = 1.01$ and $T^*/T_c^* = 1.002$ we obtain approximately $\lambda^* = 1.167$ nm, thus $\text{Kn} = 1.556 \cdot 10^{-3}$, which is still not in violation of the continuum hypothesis.

Using the thermoacoustic scaling proposed in section II.B, eq. (36) can be made dimensionless, yielding

$$\dot{Q}(x,t) = l_0^* f^*(x^*) \frac{\Omega^*(t^*)}{\rho_0^* a_0^* c_{p_0}^* / \alpha_{p_0}^*} = f(x)\Omega(t). \tag{39}$$

*III.C.2.   Explored thermodynamic parameter space for initial conditions*

| symbol | EoS | $p_0^*/p_c^*$ | $T_0^*$(K) | $a_0^*$(m/s) | $\alpha_{p_0}^*$(1/K) | $c_{p_0}^*$(kJ/(kg·K)) | $\alpha_{p_0}^* T_0^*$ |
|---|---|---|---|---|---|---|---|
| ● | PR | 1.00 | 304.12820530 | 237.33 | 3.96104 | 461.878 | 1204.66 |
| ■ | PR | 1.04 | 305.89789820 | 244.92 | 0.37584 | 45.950 | 114.97 |
| ◆ | PR | 1.40 | 319.52551062 | 272.86 | 0.03980 | 5.991 | 12.72 |
| ○ | IG | 1.00 | 608.25640530 | 385.01 | 0.0016440 | 0.8404 | 1 |
| □ | IG | 1.04 | 610.02609820 | 385.57 | 0.0016393 | 0.8404 | 1 |
| ◇ | IG | 1.40 | 623.65371062 | 389.86 | 0.0016035 | 0.8404 | 1 |

Table 2.   Initial conditions for the computational cases. The two PR and IG (ideal gas) EoS' are used.

The choice of $a_0^*$ and $\rho_0^* a_0^{*2}$ as scaling parameters for velocity and pressure fluctuations is well-established in linear acoustics, stemming from a simple inspection of the linearized isentropic equations. This choice is still appropriate for real fluids, however the authors have noticed that scaling temperature fluctuations with the base temperature $T_0^*$ only yields similarity for an ideal gas, and it is not the most general choice for a compressible fluid with a generic EoS. As shown in section II, the correct scaling parameter is $(\gamma_0 - 1)/\alpha_{p_0}^*$. To test this, we have selected initial conditions (base state) along the pseudo-boiling line.

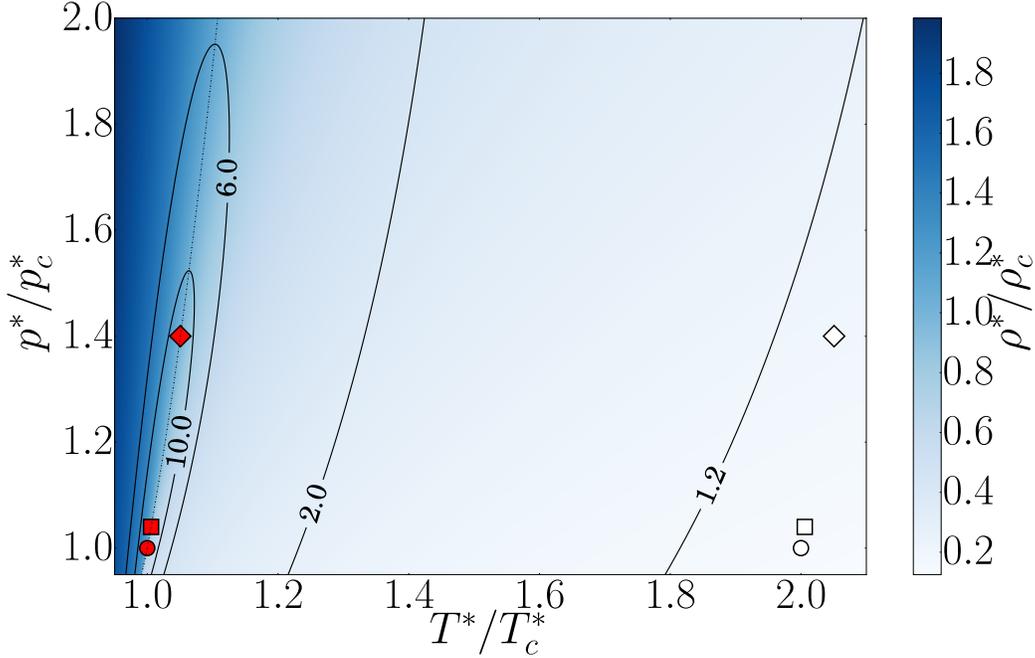

Figure 4. Flooded contours of reduced density $\rho^*/\rho_c^*$, in a reduced temperature $T^*/T_c^*$-pressure $p^*/p_c^*$ plot, with isolines of $\alpha_p^* T^*$ (solid lines). The dashed line represents the locus of maxima of $\alpha_p^* T^*$ for different pressures, here chosen as the pseudo-boiling line. The ideal gas region, on the right, is characterized by $\alpha_p^* T^* \simeq 1$. Symbols represent initial conditions used for the computations, also reported in table 2. Real-fluid and ideal-gas conditions are indicated with filled red and white symbols, respectively.

Figure 4 shows flooded contours of density, where dark and light regions correspond to heavy pseudo liquid and light pseudo gas, respectively. Those are divided by a pseudo-phase change region where the properties change rapidly but continuously, indicating a gradual transformation in the supercritical region, not abrupt as an actual phase change. Around this region the thermal expansion and the specific heat coefficients reach their maximum, defining the so-called pseudo-boiling line[1, 25] or Widom line, while the speed of sound reaches a minimum. We select the three base pressures $p_0^* = 1.00, 1.04, 1.40\ p_c^*$ on the pseudo-boiling line, thus defining the six initial base conditions described in table 2.

For each of the three selected pressures in table 2, heat injection intensities $\Omega^*$ in the range $10^3 - 10^{11}$ W/m$^2$ have been applied, bracketing the value of 1830 W/m$^2$ used by Miura et al.[21] and spanning 8 orders of magnitude.

## IV. Numerical results and discussion

The results section is divided into four parts: first, the generation of compression waves from heat release is analyzed in section IV.A with a detailed description of the phenomena of the early time stages of evolution; verification of the Rankine-Hugoniot jump conditions and a grid convergence study is then shown (section IV.B); thermoacoustically scaled shock waves are shown in section IV.C. Finally, the long term evolution is discussed in section IV.D.

### IV.A. Compression waves generation

Heat release applied locally to a compressible fluid generates a confined region of hot expanding fluid, mechanically compressing, as a piston would, the quiescent surroundings. Due to the compressibility of the fluid, propagating compressions are generated. The spatio-temporal history of the generation and early-time propagation of the compression waves from the heat release spot is shown in figure 5 for a mild heat release of $\Omega^* = 10$ kW/m$^2$. In figure 5 we can see, from left to right, the temperature rising in a quasi-isobaric post-compression state, and the antisymmetry of velocity fluctuations, together with the traveling temperature, pressure and velocity jumps characteristic of compression waves. These compressions can be approximated as acoustic waves traveling at the base speed of sound. For higher heat release rates, the post-compression state is not isobaric and the waves are shocks with measurable departures from unitary Mach numbers.

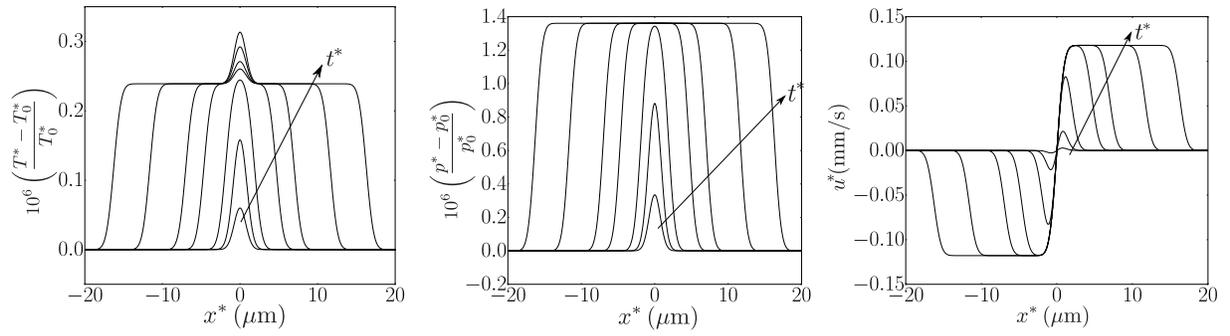

**Figure 5.** Pressure, temperature and velocity (from left to right) at times $t^* = 1, 3, 8, 20, 30, 50, 70$ ns for $p_0^*/p_c^* = 1.00$, $T_0^*/T_c^* = 1.00$ and $\Omega^* = 10$ kW/m$^2$.

### IV.B. Verification against Rankine-Hugoniot jumps and grid sensitivity analysis

*IV.B.1. Rankine-Hugoniot*

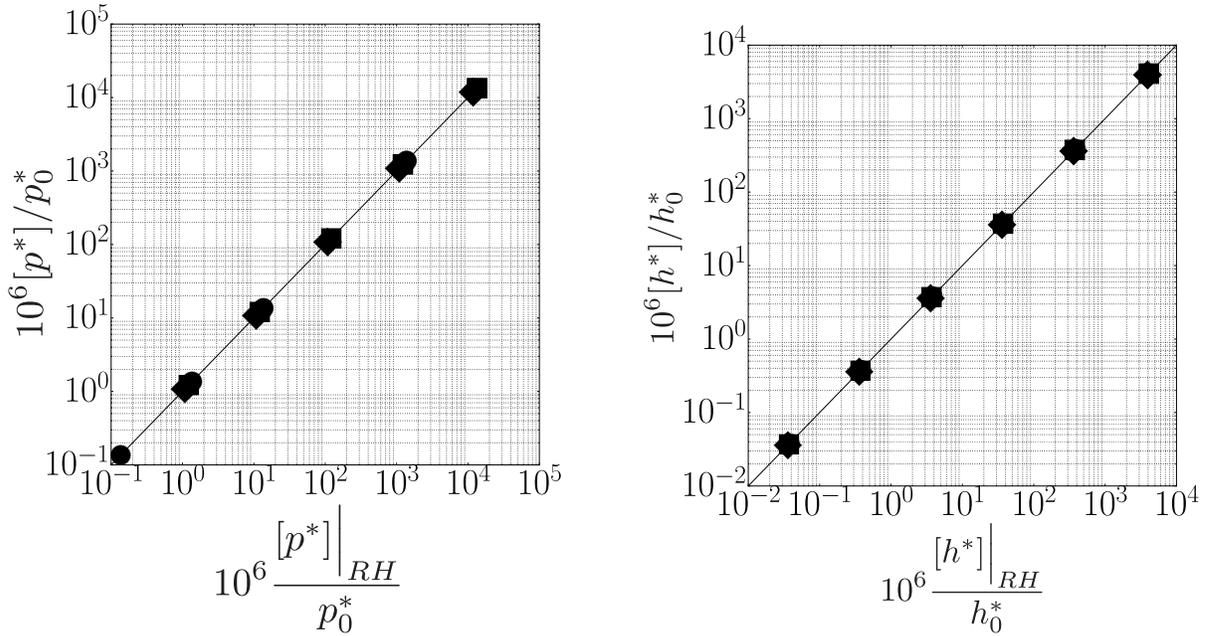

(a) Momentum RH verification.

(b) Enthalpy RH verification.

**Figure 6.** Validation for the numerical cases against Rankine-Hugoniot jump conditions.

Relative mass, momentum and total enthalpy are conserved across a steady waveform moving at constant speed $u_s^*$ according to the Rankine-Hugoniot jump conditions:

$$[\rho^* u_{rel}^*] = 0 \qquad [p^* + \rho^* u_{rel}^{*2}] = 0 \qquad \left[h^* + \frac{1}{2}u_{rel}^{*2}\right] = 0 \qquad (40)$$

where $u^*$ is the absolute (or eulerian) flow velocity and $u_{rel}^* = u^* - u_s^*$ is the flow velocity relative to the moving wave reference frame. Subscripts "0" and "1" are used to indicate variables upstream and downstream of the wave, while $[\phi^*] = \phi_1^* - \phi_0^*$ is a generic dimensional variable jump.

Conditions in eq. (40) give a nonlinear relation between upstream and downstream states of the wave. The wave speed $u_s^*$ can be computed from mass conservation, and substituted into the momentum and enthalpy conservation to obtain the post-shock values of pressure and enthalpy:

$$\begin{cases} u_s^* = (\rho_1^* u_1^* - \rho_0^* u_0^*)/(\rho_1^* - \rho_0^*) \\ p_1^*\big|_{RH} = p_0^* + \rho_0^*(u_0^* - u_s^*)^2 - \rho_1^*(u_1^* - u_s^*)^2 \\ h_1^*\big|_{RH} = h_0^* + \frac{1}{2}\left((u_0^* - u_s^*)^2 - (u_1^* - u_s^*)^2\right) \end{cases} \quad (41)$$

In order to verify that the numerical results satisfy the Rankine-Hugoniot jump conditions, post-shock values of velocity and density, $u_1^*, \rho_1^*$ have been extracted from the numerical simulations and inserted in eq. (41), returning the expected pressure and enthalpy in the post-shock state. The latter are then compared with the simulation data (figure 6) confirming the sanity of Navier-Stokes computations. This verification, carried out in post-processing, does not require explicit calls to the PR EoS.

### IV.B.2. Grid convergence

Nondimensional variations of temperature and pressure, and velocity profiles are shown in figure 7 for the different grid resolutions described in table 3, showing oscillations for coarse ones.

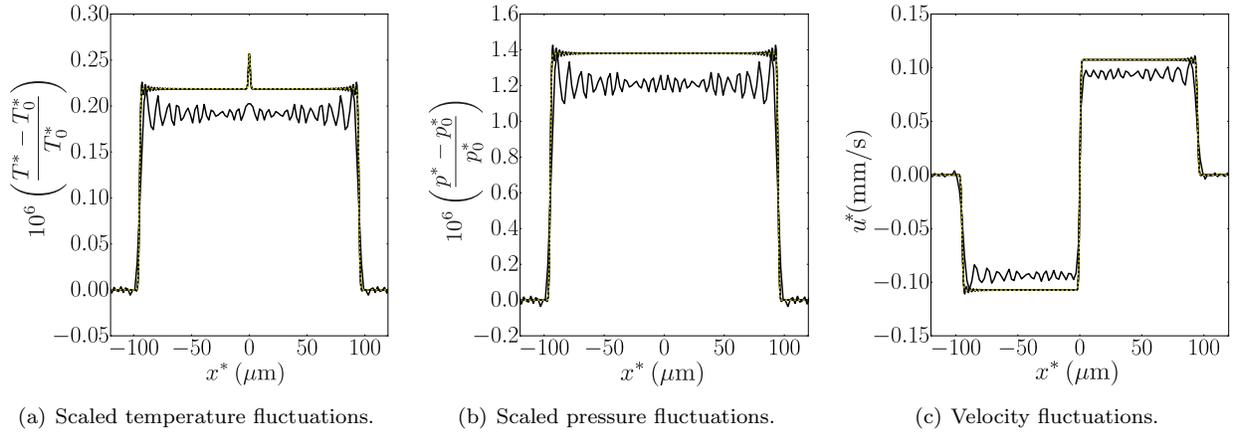

(a) Scaled temperature fluctuations.  (b) Scaled pressure fluctuations.  (c) Velocity fluctuations.

**Figure 7.** Spatial profiles at $t = 0.4$ $\mu$s for $p_0^*/p_c^* = 1.00$, $T_0^*/T_c^* = 1.00$, and $\Omega^* = 10$ kW/m$^2$, with grids defined in table 3. Higher values of $N$ correspond to smoother waveforms.

| $N$ | $L^*$(mm) | $\Delta x^*$ (nm) | $CFL$ | $\Delta t^*$ (ns) |
|---|---|---|---|---|
| 1000 | 2 | 2000 | 0.1 | 5.30e-1 |
| 4000 | 2 | 500 | 0.1 | 1.33e-1 |
| 16000 | 2 | 125 | 0.1 | 3.31e-2 |
| 64000 | 2 | 31.25 | 0.1 | 8.29e-3 |
| 128000 | 2 | 15.625 | 0.1 | 4.14e-3 |

**Table 3.** Grid parameters for grid convergence study. A duct longer than the one described in section III.C is here considered.

Table 3 shows a progressively refined uniform grid, employing a larger number of grid points $N$, fixing duct length $L^*$ and CFL so that $\Delta x^*$ and $\Delta t^*$ vary accordingly. The reported $\Delta t^*$ is averaged over the first 100 iterations.

The finest grid is chosen as reference for the RMS error computations. The RMS error of the profile shown in figure 7a is plotted in figure 8 showing an effective third order decay rate of the error.

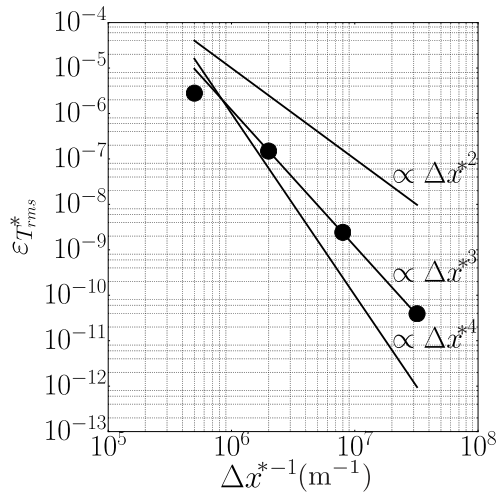

Figure 8. Temperature RMS error for for grid resolutions in table 3 and for the case shown in figure 7a.

### IV.C. Shock generation and thermoacoustic scaling

The phenomenology described in section IV.A qualitatively holds in all cases however, depending on the value of $\Omega^*$, heat-release-generated compressions can be weak, mild or strong shock waves. As shown in figure 9, in fact, the speed of the compressions $u_s^*$ is higher than the local speed of sound in the case of shocks, while the relative Mach number of isentropic waves is slightly higher than unity. Strong thermoacoustic shock waves reveal additional features such as a non-uniform post-shock state, caused by the high thermodynamic nonlinearities.

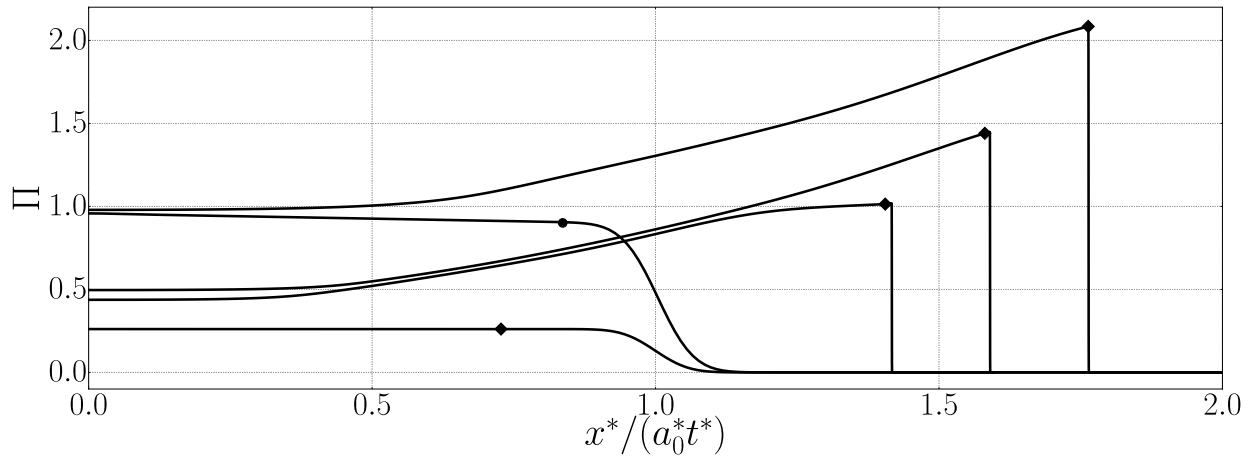

Figure 9. Shock strength $\Pi$ versus scaled spatial coordinate $x^*/(a_0^* t^*)$ equal to the propagation Mach number, relative to state 0. Plots are for $\Omega^* = 10^5, 10^8, 3 \cdot 10^{10}, 6 \cdot 10^{10}, 10^{11}$ W/m² at times $60, 60, 46, 46, 25$ ns. The symbols indicate the initial condition and the value chosen as post-shock state. The values of $\Pi$ were multiplied by $10^5, 200, 2, 2, 2$, respectively, to make them fit in a single figure.

The amplitude of the heat-release-induced isentropic thermoacoustic compressions can be predicted from an existing relationship by Miura et al.[21] which has been recast here into

$$[p^*] = \frac{1}{2} \frac{a_0^* \alpha_{p_0}^*}{c_{p_0}^*} \Omega^* \qquad (42)$$

using the thermodynamic relation $(\partial T^*/\partial p^*)_{s,0} = T_0^* \alpha_{p_0}^*/(\rho_0^* c_{p_0}^*)$. Eq. (42) can be made dimensionless with our approach yielding the very elegant relationship

$$\Pi = \Omega, \qquad (43)$$

between the dimensionless pressure jump (shock strength[31]),

$$\Pi = \frac{[p^*]}{\rho_0^* a_0^{*2}}, \tag{44}$$

and the dimensionless heat release rate

$$\Omega = \frac{\Omega^*/2}{\rho_0^* a_0^* c_{p_0}^*/\alpha_{p_0}^*}. \tag{45}$$

Eq. (43) allows computation of all the dimensionless jumps together with other relevant quantities, such as the acoustic flux $W^*_{acou} = [p^*][u^*]$, which reads $W_{acou} = \Pi^2$ in dimensionless form for purely traveling waves.

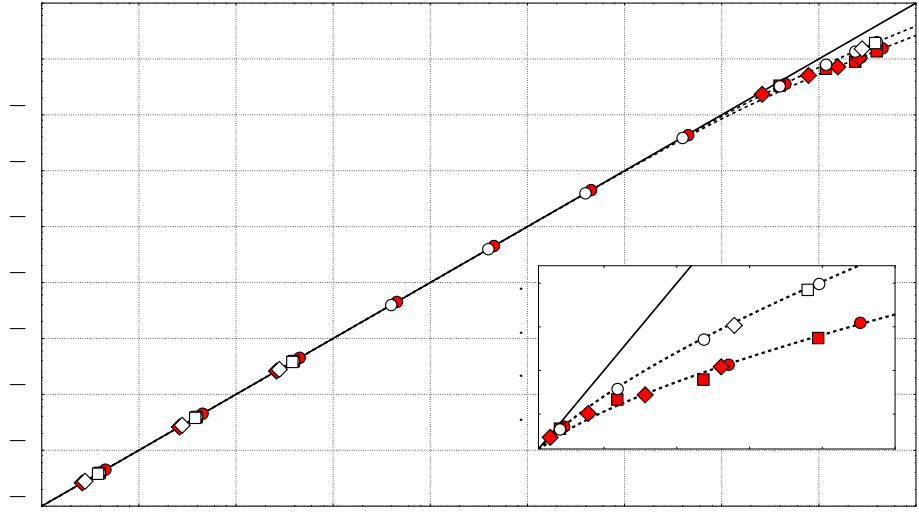

Figure 10. Shock strength versus nondimensional heat release from numerical computations and models. Symbols represent data from the numerical simulations with $\Omega^*$ in the range $10^3 - 10^{11}$ W/m$^2$. The solid line is the isentropic prediction $\Pi = \Omega$. The dashed lines are obtained from eq. (46) capturing both isentropic and non-isentropic compressions (shock waves) for both ideal and real gas EoS. The dashed line crossing ideal gas cases (white-filled symbols) is obtained for $\gamma_0 = \widetilde{\gamma}_0 = 1.29$, while the one crossing real-fluid data (red-filled symbols) is obtained for $\widetilde{\gamma}_0 = 3.36$.

The intensity of these heat-release-generated shocks can be predicted by extending the parametrization initially derived by Chu[6] for inviscid ideal gases to a general fluid,

$$\Omega = \sqrt{2}\frac{(\widetilde{\gamma}_0\Pi + 1)\Pi}{\sqrt{(\widetilde{\gamma}_0 + 1)\Pi + 2}}, \tag{46}$$

via the isentropic exponent

$$\widetilde{\gamma}_0 = \frac{a_0^{*2}\rho_0^*}{p_0^*} \tag{47}$$

acting like an "effective ratio of specific heats".[14] Notice that in the limit of $\Pi \to 0$ eq. (46) reverts to eq. (43). This confirms that for isentropic compressions only one degree of thermodynamic freedom is allowed, while for non-isentropic waves a second thermodynamic parameter needs to be specified. Eq. (46) can therefore predict heat-release-induced shock amplitudes in supercritical fluids, by simply extending a model initially derived for ideal gases, also aided by the proposed thermoacoustic scaling. This is illustrated in figure 10, where predictions together with results from numerical simulations show excellent matching for all cases.

### IV.D. Long term evolution: resonating shock wave regime

The heat-release-induced compression waves propagate in the enclosure and are reflected off the hard ends and thus interact with each other. As this process continues, the enclosure can be heated and pressurized at a rate dictated by the resonance frequency $u_s^*/L^*$. Heating of an enclosure can then be achieved more rapidly than in a purely conductive process, that is very slow especially in near-critical fluids. This phenomenon has been called Piston Effect (PE). In the low amplitude regime, the post-(weak) shock states are linearly superimposed, whereas for higher amplitudes more complex interactions occur. In all cases, consecutive compression waves increase mean pressure and temperature in the duct, as long as they do not vanish. As soon as the heating is halted, two expansions are generated, which cancel the effect of the compressions (negative interference) but not the waves themselves, resulting in a ceasing of the mean heating and an oscillation, about a steady non-uniform base state, of temperature and pressure fluctuations. This leaves thermal diffusion as the only thermal equilibration process left. This phenomenology has been observed in the experiments of Miura et al.,[21] where density fluctuations $\delta \rho^* \approx 10^{-7}$ g/cm$^3$ were measured with a very precise interferometer, confirmed by low-order numerical models employing asymptotic expansions.[4]

We reproduce here the experiment of Miura et al.[21] scaling the geometry down by a factor of 100, and employing isothermal or adiabatic boundary conditions (b.c.s). The initial base state is chosen so that the dimensionless intensity of the first compression from the experiments is matched by the simulations: from eq. (43) the equivalent heat release rate $\Omega^*$ can be computed. The dimensionless density variations, for both the experiment and the numerical simulations, are shown in figure 11 against dimensionless time.

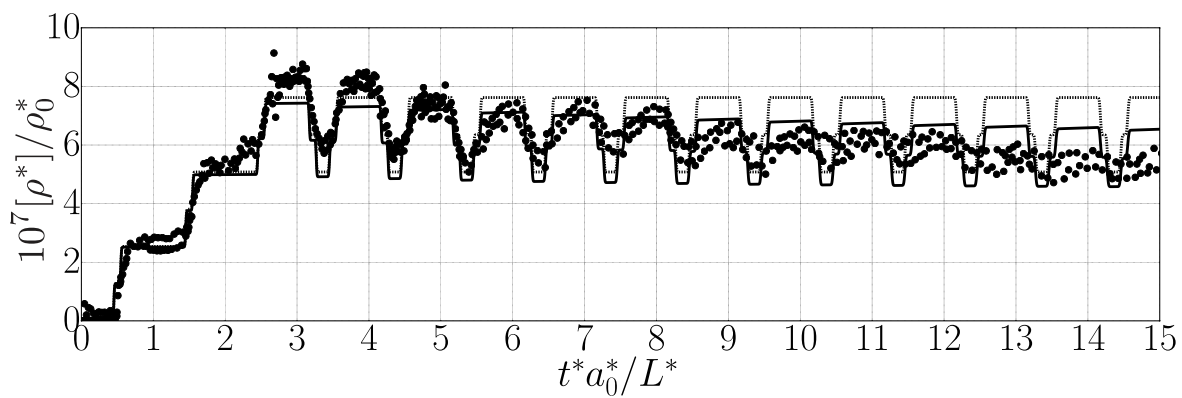

Figure 11. Dimensionless density variation with dimensionless time at the center of the enclosure showing data from the experiment of Miura et al. (full circles), and simulations with isothermal boundary conditions (solid line) and adiabatic boundary conditions (dashed line).

Very good matching between experimental result and computations is observed at the early stages. Nevertheless, as soon as the heat is removed, and the first expansion is generated, the numerical simulations begin to deviate from the experimental results. In the long term evolution, a decrease in wave amplitude appears both in the experiment and in the numerical result with isothermal boundaries. On the other hand, in the case of adiabatic walls, the fluctuation energy is conserved and therefore no decrease in amplitude occurs, if not numerical or because of dissipative effects. In the experiments, the decay rate is much larger than the one observed in numerical simulations with isothermal b.c.s. This has already been noticed in previous works[5] and still requires a complete explanation. Additional losses such as bulk viscosity effects (not accounted for in this work) are also expected to contribute to this decay. Also, as already observed by others,[12] the presence of the side walls in the experiment could be an additional source of dissipation. Nevertheless, the major source of wave energy loss are the reflections from the walls which include effects not commonly accounted for in Navier-Stokes models. For example, the very high fluid effusivity of near-critical fluids could make an isothermal b.c. invalid for modeling.[35] We in fact speculate that the solid wall imparts a finite amplitude complex impedance (to be investigated in future work) to the impinging wave, in the near-critical fluid, therefore reducing its intensity at every reflection.

## Acknowledgments

This work has been performed under the support of the Andrews and Rolls-Royce Doctoral Fellowships at Purdue University. Computing resources were provided by the Rosen Center for Advanced Computing (RCAC) at Purdue University and Information Technology at Purdue (ITaP). The authors are thankful to Jean-Pierre Hickey, Iman Rahbari, and Prateek Gupta for fruitful discussions.